\documentclass[oldversion]{aa}
\usepackage{graphicx}
\usepackage{natbib}
\setcitestyle{aysep={}} 
\usepackage{hyperref}
\usepackage{longtable}
\usepackage{lscape}
\usepackage{color}

\begin{document}

   \title{The B-Star Exoplanet Abundance Study: a co-moving 16-25 $M_{\rm Jup}$ companion to the young binary system HIP 79098
\thanks{Based on archival observations from the European Southern Observatory, Chile (Programs 073.D-0534 and 095.C-0755).}
}


   \author{Markus Janson\inst{1} \and
          Ruben Asensio-Torres\inst{1} \and
          Damien Andr{\'e}\inst{1} \and
          Micka{\"e}l Bonnefoy\inst{2} \and
          Philippe Delorme\inst{2} \and
          Sabine Reffert\inst{3} \and
          Silvano Desidera\inst{4} \and
          Maud Langlois\inst{5} \and
          Ga{\"e}l Chauvin\inst{6,2} \and
          Raffaele Gratton\inst{4} \and
          Alexander J. Bohn\inst{7} \and
          Simon C. Eriksson\inst{1} \and
          Gabriel-Dominique Marleau\inst{8} \and
          Eric E. Mamajek\inst{9,10} \and
          Arthur Vigan\inst{11} \and
          Joseph C. Carson\inst{12}
          }

   \institute{Department of Astronomy, Stockholm University, Stockholm, Sweden\\
              \email{markus.janson@astro.su.se}
         \and
            Univ. Grenoble Alpes, IPAG, Grenoble, France
        \and
            Landessternwarte, Zentrum f{\"u}r Astronomie der Universit{\"a}t Heidelberg, Heidelberg, Germany
        \and
            INAF - Osservatorio Astronomico di Padova, Padova, Italy
        \and
            CRAL, CNRS, Universite Lyon, Saint Genis Laval, France
        \and
            Unidad Mixta Internacional Franco-Chilena de Astronom\'{i}a, CNRS/INSU and Departamento de Astronom\'{i}a, Universidad de Chile, Santiago, Chile
        \and
            Leiden Observatory, Leiden University, Leiden, The Netherlands
        \and
            Institut f{\"u}r Astronomie und Astrophysik, Eberhard Karls Universit{\"a}t T{\"u}bingen, T{\"u}bingen, Germany
        \and
            Jet Propulsion Laboratory, California Institute of Technology, Pasadena, CA, USA
        \and
            Department of Physics and Astronomy, University of Rochester, Rochester, NY, USA
        \and
            Aix Marseille Universit{\'e}, CNRS, LAM, Marseille, France
        \and
            College of Charleston, Charleston, SC, USA
             }

   \date{Received ---; accepted ---}

   \abstract{Wide low-mass substellar companions are known to be very rare among low-mass stars, but appear to become increasingly common with increasing stellar mass. However, B-type stars, which are the most massive stars within $\sim$150 pc of the Sun, have not yet been examined to the same extent as AFGKM-type stars in that regard. In order to address this issue, we  launched the ongoing B-star Exoplanet Abundance Study (BEAST) to examine the frequency and properties of planets, brown dwarfs, and disks around B-type stars in the Scorpius-Centaurus (Sco-Cen) association;   we also analyzed archival data of B-type stars in Sco-Cen. During this process, we identified a candidate substellar companion to the B9-type spectroscopic binary HIP 79098 AB, which we refer to as HIP 79098 (AB)b. The candidate had been previously reported in the literature, but was classified as a background contaminant on the basis of its peculiar colors. Here we demonstrate that the colors of HIP 79098 (AB)b are  consistent with several recently discovered young and low-mass brown dwarfs, including other companions to stars in Sco-Cen. Furthermore, we show unambiguous common proper motion over a 15-year baseline, robustly identifying HIP 79098 (AB)b as a bona fide substellar circumbinary companion at a 345$\pm$6 AU projected separation to the B9-type stellar pair. With a model-dependent mass of 16-25 $M_{\rm Jup}$ yielding a mass ratio of $<$1\%, HIP 79098 (AB)b joins a growing number of substellar companions with planet-like mass ratios around massive stars. Our observations underline the importance of common proper motion analysis in the identification of physical companionship, and imply that additional companions could potentially remain hidden in the archives of purely photometric surveys.}


\titlerunning{BEAST I: HIP 79098 (AB)b}
\authorrunning{M. Janson et al.}

   \maketitle
%

\section{Introduction}
\label{s:intro}

As the field of high-contrast imaging develops, it is revealing an increasing number of massive planets and low-mass brown dwarf companions, primarily around stars more massive than the Sun  \citep[e.g.,][]{marois2008,lagrange2010,carson2013,macintosh2015}. Direct imaging is particularly  suitable for studying young systems since the brightness contrast between the primary star and a substellar companion is  minimized when the planet is newly formed. At an age in the range of 10-20~Myr \citep{pecaut2016} and a distance of 120-150~pc \citep{brown2018}, Scorpius-Centurus \citep[Sco-Cen;][]{dezeeuw1999} is the nearest large young stellar region, and has therefore been particularly fruitful source of  such companions \citep[e.g.,][]{lafreniere2009,rameau2013,bailey2014,chauvin2017,cheetham2018,keppler2018}. We  recently launched the B-star Exoplanet Abundance Study (BEAST), which is an ESO\footnote{European Southern Observatories} Large Program dedicated to the study of planetary companions around the most massive stars in Sco-Cen. BEAST will be observing 83 B-type members of Sco-Cen with SPHERE \citep{beuzit2019} at the VLT\footnote{Very Large Telescope}. The observations will reveal whether the frequency of massive giant planets continue to  increase with stellar mass, or whether there is a turnover somewhere along the B-type range, signifying an optimal stellar mass for planet formation. 

In the target selection process for BEAST, we removed targets that had been previously observed with SPHERE in order to avoid unnecessary target duplications. However,  to maintain completeness for the survey, we are also continually analyzing the archival data to evaluate their detection space and point-source candidates. In this process, the HIP 79098 (HR 6003, HD 144844) system has proven to be a particularly interesting system. As we  see in Sect. \ref{s:binary}, HIP 79098 is a B9-type member of Upper Scorpius and consists of a close stellar spectroscopic binary. We  refer to the stellar components as HIP 79098 A and B, and thus the central unresolved pair as HIP 79098 AB. We identified a candidate substellar companion to HIP 79098 AB in archival SPHERE data which, as we  show in the following, closely shares a common proper motion with the central stellar pair. We  refer to it  as HIP 79098 (AB)b.

The candidate companion was first noticed by \citet[][hereafter ST02]{shatsky2002}. They detected a large number of point sources in their ADONIS coronagraphic imaging data around massive stars, and distinguished physical binary pairs from optical pairs on the basis of photometric matching to stellar isochronal models. Since HIP 79098 (AB)b was too faint and too red to match those models, ST02 classified it as a probable reddened background star. It is therefore listed as an optical (i.e., non-physical) component in their tables. At the  time of writing\footnote{We refer here to the online version of the catalog as accessed through the VIZIER service, which is continuously updated to account for new binarity information. Checked in April 2019.}, the Washington Double Star catalog \citep[][]{mason2001} also lists the point source as non-physical based on the ST02 results under ID SHT61, although the individual note for the target mistakenly labels the classification as being based on proper motion analysis. Subsequent to the ST02 result, HIP 79098 (AB)b was independently detected by \citet[][hereafter K05]{kouwenhoven2005}. The data were also acquired with ADONIS at a similar time, with observations made in 2000 and 2001 for the survey. Their  photometric classification of candidates  was similar to that of  ST02. On this basis they also classified it as a background star in their tables, although they noted in the text that it cannot be formally excluded that the companion could be physically bound. The same investigators re-observed HIP 79098 (along with several other targets) with NACO\footnote{NAOS-CONICA \citep{lenzen2003,rousset2003}.} in 2004  \citep[][hereafter K07]{kouwenhoven2007}, and  performed a more detailed photometric check and concluded a background status for HIP 79098 (AB)b. Neither ST02 nor K05 or K07 performed any common proper motion (CPM) analyses to test whether the system is bound on an astrometric basis.

As in the earlier studies, the K07 conclusion is based on the fact that HIP 79098 (AB)b is significantly redder than  would be expected from conventional isochronal models, in this case represented by \citet{chabrier2000}. However, as the study of substellar objects has progressed over the past decade, we now know that substellar objects display a wide range of photometric properties, which cannot all be represented by a single set of one-parameter models. In particular, it is known that young substellar objects generally display considerably redder colors than their old field counterparts of the same spectral type \citep[e.g.,][]{liu2013,gizis2015,bonnefoy2016,faherty2016}, probably due to their lower surface gravities. Thus, there is a substantial risk of systematic misclassification when applying conventional isochronal models to young objects such as HIP 79098 (AB)b and other potential Sco-Cen members. For this reason, CPM analysis is considered a more reliable (and model-independent) method for testing physical companionship, and is the standard means of assessment for candidates in contemporary direct imaging surveys \citep[e.g.,][]{brandt2014,vigan2017,asensio2018}. In this paper, we re-examine the literature data along with additional archival data for astrometric as well as photometric analysis in order to update the status of the companion HIP 79098 (AB)b, and discuss the implications of the results for wide candidate companions in the literature.

\section{Properties of the host system}
\label{s:binary}

The host system HIP 79098 has an unresolved spectral type (SpT) which is generally classified as B9, with more detailed classifications including  B9IVn+Ap(Si)s \citep{abt1995}. According to the \textit{Gaia} DR2 parallax \citep{brown2018}, the system distance is 146.3$\pm$2.5\,pc. The color excess $E(B-V)$ for HIP 79098 in the literature is 0.12$\pm$0.02 mag \citep[e.g.,][]{norris1971,castelli1991,pecaut2013,huber2016}. Following \citet{fiorucci2003}, this gives a visual extinction of $A_V = 0.38 \pm 0.06$ mag, which in turn gives an absolute magnitude of $M_V = -0.33 \pm 0.07$ mag.

HIP 79098 is a member of Upper Scorpius (USco), which is the youngest subregion of Sco-Cen, thus implying an age of 10$\pm$3 Myr \citep{pecaut2016}. The USco membership has been under consideration for a long time \citet{bertiau1958,dezeeuw1999}, and is supported with contemporary Bayesian membership tools such as BANYAN $\Sigma$ \citep{gagne2018}, which gives a 98\% probability that HIP 79098 is a member of USco based on Gaia DR2 astrometry. Radial velocity (RV) was not used in the BANYAN $\Sigma$ analysis due to the reported spectroscopic binarity of HIP 79098. The identification of HIP 79098 as a spectroscopic binary is based on strong radial velocity (RV) variability \citep[e.g.,][]{levato1987,worley2012}. If we assume that the unresolved SpT of the system reflects the SpT of the HIP 79098 A component, this SpT implies a primary stellar mass of approximately 2.5~$M_{\rm sun}$ at the age of USco \citep{lafreniere2014}. Some sources report double lines in the spectrum \citep{hartoog1977,schneider1981,brown1997}, which implies that the B component is probably also quite massive. \citet{norris1971} reports a flux difference of approximately a factor of 3 between the primary and secondary based on spectroscopic data. A massive secondary is supported by the RV variability of the primary line, which spans from -42\,km/s to 73\,km/s among the 12 epochs from \citet{levato1987} and \citet{worley2012}. Meanwhile, \citet{becker2015} cite an RV of -16.95$\pm$1.87 from HIRES data over a time span of 29 days with no mention of  double lines.

It is not clear that the different RV related measurements in the literature can provide a homogenous picture of the central binary. On the one hand, \citet{worley2012} implies strong RV variability even on single-day timescales (e.g., 73 km/s on MJD 53900 versus 37 km/s on MJD 53901);  on the other hand, -16.95$\pm$1.87 km/s over 29 days in 13 separate spectra from \citet{becker2015} implies much slower (if any) motion. A highly eccentric orbit could in principle accommodate  similar variations. In fact, we can fit both the \citet{levato1987} and \citet{worley2012} RVs simultaneously with a 558.5 day, $e=0.88$ orbit and an amplitude of 148 km/s. However, this combination of period and amplitude gives an unreasonably high minimum mass for the secondary of 24~$M_{\rm sun}$. This would correspond to an O-type star rather than the system SpT of B9. Even under the assumption of a single star giving rise to all the flux from HIP 79098, its $M_V$ of -0.33 mag is inconsistent with any SpT earlier than B5, which corresponds to a mass of $\sim$4.2 $M_{\rm sun}$, much lower than the 24~$M_{\rm sun}$ that would be required. The mass of the B component in the RV fit can be substantially decreased if, for example,  we allow for  an additional linear trend in the fitting, but this would require a third stellar component (probably with an unreasonable mass itself), or some large systematic offset between the different data sets. Fully determining the true parameters of the central binary will therefore be a complicated task that stretches beyond the scope of this paper. Here we simply note that the total mass of the system should range somewhere from 2.5~$M_{\rm sun}$ if the mass is dominated by HIP 79098 A, up to 5~$M_{\rm sun}$ if the binary consists of a nearly equal-mass pair of late B-type stars. A future dedicated study could plausibly provide significantly tighter constraints on the component masses and other system parameters.

Given that the \textit{Gaia} and \textit{Hipparcos} proper motions only differ by 2.4 mas/yr, we do not expect photocenter motion of the central binary to affect the relative astrometric analysis of HIP 79098 (AB)b in Sect. \ref{s:astro}. This is further supported by the fact that most lines of evidence point to a small orbit for HIP 79098 AB, and  that HIP 79098 AB shows no deviation from a point-like morphology in the unsaturated images taken for the photometric calibration discussed in Sect. \ref{s:photo}.

We also note that there is a low-mass star at 65$^{\prime \prime}$ (9500 AU) separation, designated 2MASS J16084836-2341209 (abbreviated here as J1608), whose proper motion is quite similar to that of  HIP 79098. J1608 was independently identified as an USco member by \citet{lodieu2007}. It was characterized as an M5-type star in \citet{lodieu2011}, which they translated to an isochronal \citep{baraffe1998} mass of 0.12~$M_{\rm sun}$ based on an age of 5 Myr for USco \citep{preibisch2002}. With our older adopted age estimate of 10 Myr, the corresponding mass becomes 0.16~$M_{\rm sun}$. J1608 has also been flagged as disk-bearing in \citet{riaz2012,luhman2012}, which further supports a young age. 

The \textit{Gaia} DR2 parallaxes of HIP 79098 and J1608 are consistent to better than 2$\sigma$, while their proper motions differ in RA by about 2 mas/yr. This difference is formally significant at nearly 8$\sigma$; however, the binarity of HIP 79098 could lead to an underestimation of its error, as also indicated by the fact that it is flagged for excess noise in \textit{Gaia}. From the available data, we cannot conclude whether J1608 is a very wide companion to HIP 79098, or whether it is a separate low-mass member of the Sco-Cen association. 

Similarly, there is another low-mass star at 88$^{\prime \prime}$ (12900 AU) separation with the designation 2MASS J16083908-2340055 (hereafter J160839). J160839 was identified as being part of USco in \citet{luhman2012} and assigned an M5 SpT classification. Unlike J160848, J160839 shows no evidence for infrared excess in \citet{luhman2012}. However, it is shown to exhibit a peculiar short-period ($\sim$0.7 days) variability in \citet{stauffer2018}. The variability pattern has similarities to the ``scallop-shell'' variability discussed in \citet{stauffer2018}, which is a class of variability seen only in young stellar populations.

The parallax of J160839 is fully consistent with HIP 79098, differing only by 0.5$\sigma$. However, the proper motion differs by nearly 15$\sigma$. As mentioned previously, a direct comparison of proper motion to this degree of precision is compromised by the multiplicity of HIP 79098. In addition, the variability analysis in \citet{stauffer2018} implies that J160839 might  be binary, which would further complicate the proper motion analysis. In addition to relating J160848 and J160839 to HIP 79098 individually, we can also compare the two low-mass stars to each other. Their parallaxes are quite similar, with only a 1.5$\sigma$ difference, while their proper motions differ by about 8$\sigma$. The conclusion for J160839 is therefore the same as for J160848: there is an intriguing possibility of companionship with HIP 79098, but more data will be required to test this scenario. The possibility of one or two additional low-mass objects at very wide separation adds further interest regarding the study of the architecture of the system, and may potentially provide clues on its history. 

\section{Data acquisition and reduction}
\label{s:obs}

We have identified several  archival or literature data sets where the companion is visible: (1) a set from ADONIS in 2000 in the $J$ and $K_{\rm s}$ bands, originally published in ST02; (2) NACO $J$, $H$, and $K_{\rm s}$ data from 2004 published in K07; and (3) a previously unpublished SPHERE data set from 2015 in $K_1$ and $K_2$. There is also the $K_{\rm s}$ ADONIS data set presented in K05, consistent with ST02 and approximately contemporaneous but with a less precisely specified time stamp, and we thus omit it in this analysis. 

The ST02 data are not archived, but the full survey data were acquired during the nights of \textcolor[rgb]{0.984314,0.00392157,0.0235294}{May 24-28, 2000}, so HIP 79098 must have been observed in that range of dates. Integration times for the coronagraphic observations were 3-5 s per frame, with 20 on-source frames per pointing, and HIP 79098 was observed in four coronagraphic pointings. This means that the total integration time was in the range of 320$\pm$80~s. The NACO data set is available in the ESO archive, and consists of three photometric bands with identical settings taken on 9 Jun 2004. Three on-target frames (and three sky frames) per band were acquired, each with 35 subintegrations of 0.35~s, giving a total on-source integration time of 37~s. The primary star is saturated in those images, but a sequence of images with a neutral density (ND) filter were also acquired, allowing for non-saturated imaging of the primary and thus enabling photometric calibration for the companion. Likewise, the SPHERE data set is available in the ESO archive. The companion is too far away to be included in the IFS field of view (FOV), but it is comfortably encompassed by the IRDIS FOV. The set contains 16 frames of $4 \times 16$~s each, for a total integration time of 1024~s. While the observing sequences are short, the companion is bright enough to be visible in individual raw frames, so its properties can still be well determined in the existing data.

\begin{figure}[htb]
\centering
\includegraphics[width=8cm]{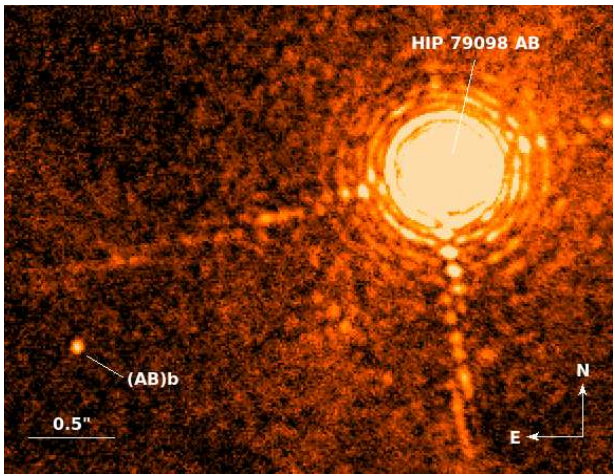}
\caption{$K_1$ image of HIP 79098 AB and its faint companion HIP 79098 (AB)b from SPHERE, without PSF subtraction.}
\label{f:hip79098im}
\end{figure}

We reduced the SPHERE data with the \textit{SpeCal} pipeline \citep{galicher2018} within the SPHERE Data Center \citep{delorme2017} framework. The field rotation during the observation was $<$1~deg, so angular differential imaging (ADI) cannot be efficiently used. Instead, we performed radial profile subtraction to eliminate the bulk of the residual stellar halo. The photometry and astrometry of the companion were then extracted through template fitting \citep{galicher2018}. An image of the system (before profile subtraction) is shown in Fig. \ref{f:hip79098im}

In order to double-check the astrometric and photometric values from K07 for the NACO data, we  also downloaded the corresponding archival data and reduced them. For this purpose, we used a fully custom pipeline to create dark and flat frames and applied them to the scientific data, subtracted a median background from the jittered data, shifted the frames to a common reference frame, subtracted a radial average PSF profile of the primary star, and median combined the frames. The same steps, except for the radial profile subtraction, were applied to the ND filtered images of the system. Registration of the primary was done by using a Moffat profile to fit the wings of the PSF since the core was mildly saturated in the non-ND frames. The secondary in the non-ND frames and the primary in the ND frames could be fit with a Gaussian profile. For pixel scale and true north orientation, we adopted values of 13.23$\pm$0.05 mas/pixel and 0.14$\pm$0.25 deg respectively, based on NACO calibrations for NACO 2004 data as presented in \citet{neuhauser2005}. We selected the $K_{\rm s}$ band for astrometry since it has the highest $S/N$ for the companion (the astrometric values in $J$ and $H$ are consistent within the error bars).

Aperture photometry was performed with a range of different aperture sizes up to a radius of 3 pixels. This is particularly important for the $J$ band where the companion is very faint and sensitive to the exact background level. We get consistent results using apertures of  different sizes,  to within 0.04~mag, which is a much smaller scatter than the dominating noise discussed below. While the NACO manual states a typical transmission value of 1/80 for the ND filter that is used for the $JHK$ bands, the actual transmission varies slightly from filter to filter. Thus, to acquire more precise photometric calibration, we read out the transmission curve of the ND filter (also available in the NACO manual) at the central wavelengths of the respective bands. As a result, we derive transmission factors of 1.36\% in $J$, 1.38\% in $H$, and 1.43\% in $K_{\rm s}$. The dominating error in the photometry arises from the rather unstable ambient conditions, which give rise to a considerably larger scatter than  would be present under photon noise-limited conditions.

\section{Astrometric analysis}
\label{s:astro}

\begin{figure}[htb]
\centering
\includegraphics[width=8cm]{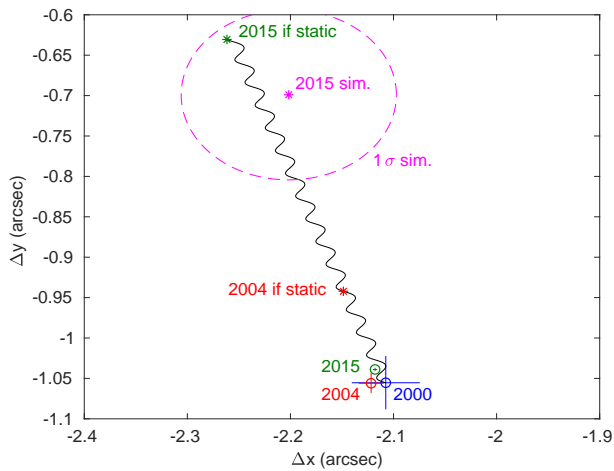}
\caption{Proper motion analysis of HIP 79098 (AB)b. The blue, green, and red circles with error bars are the measured positions of the companion relative to the parent star for epochs 2000, 2004, and 2015 respectively. Each observation is consistent with CPM. Also plotted is a static background track in black starting from the 2000 epoch, with red and green asterisks denoting the expected locations for a static background object in 2004 and 2015. The magenta asterisk and dashed line are the mean and 1$\sigma$ error ellipse of the simulated sample of galactic stars (see text).}
\label{f:astro}
\end{figure}

The astrometric values that we  derived, along with the literature astrometry from ST02, are listed in Table \ref{t:astrometry}. They are plotted along with the prediction for a static background object in Fig. \ref{f:astro}. All epochs of observation are fully consistent with CPM, and clearly distinct from the static background hypothesis. It is important to note, however, that being distinct from the static trajectory does not, by itself, prove the CPM hypothesis. Background objects  have some degree of proper motion, and for a target with a relatively low proper motion, such as stars in the Sco-Cen region, the magnitudes of the proper motion can occasionally be comparable. This has been demonstrated for the case of HD 131399 Ab, originally thought to be a CPM object \citep{wagner2016}, but later shown to display a distinct proper motion from the primary star by an amount exceeding the expected escape velocity \citep{nielsen2017}. Based on the new proper motion analysis and on spectroscopic analysis, \citet{nielsen2017} concluded that the candidate was more likely to be an unusual background contaminant, unrelated to the primary star. 

\begin{table}[htb]
\caption{Astrometric data of HIP 79098 B.}
\label{t:astrometry}
\centering
\begin{tabular}{lllll}
\hline
\hline
Date & MJD & Facility & Sep & PA \\
 & (d) & & ($^{\prime \prime}$) & (deg) \\
\hline
2000-05-26$^a$      & 51690 & ADONIS & 2.357$\pm$0.033 & 116.6$\pm$0.8 \\
2004-06-09      & 53165 & NACO & 2.370$\pm$0.011 & 116.46$\pm$0.30 \\
2015-07-20      & 57223 & SPHERE & 2.359$\pm$0.001 & 116.13$\pm$0.06 \\
\hline
\end{tabular}
\begin{list}{}{}
\item[$^{\mathrm{a}}$] Mean date for the range given in ST02, see text.
\end{list}
\end{table}

We  therefore performed a similar analysis to that in \citet{nielsen2017} to assess the hypothesis that HIP 79098 (AB)b could be a rare background contaminant. We did this using Besan\c{c}on models \citep{robin2003} generated from an online interface\footnote{https://model.obs-besancon.fr}. We generated a simulated stellar population centered on the coordinates of HIP 79098 in a 1 deg$^2$ field. A population out to 50 kpc was simulated, including all stars within 2$\sigma$ of the $K$-band brightness of HIP 79098 (AB)b, which are equivalent settings to those used in \citet{nielsen2017}. The resulting yield is a sample of 1675 stars, with a mean proper motion of $\mu_{\rm RA} = -3.96$\,mas/yr and $\mu_{\rm Dec} = -4.52$\,mas/yr, and consistent standard deviations of 6.89\,mas/yr in the RA direction and 6.95\,mas/yr in the Dec direction. This result is plotted in Fig. \ref{f:astro} for a 2000-2015 baseline, along with the static background expectation. The simulated background population is separated from the CPM location by 3.5$\sigma$. It should be noted that since the simulated population distribution is not Gaussian, this value cannot be translated into a conventional $<0.1$\% probability. The fraction of simulated stars that exceed a 3.5$\sigma$ deviation (in any direction) constitute approximately 2.6\%. Nonetheless, this analysis provides strong support for CPM, particularly since the candidate companion deviates significantly from the background locus, and is in fact also located specifically at the CPM position. 

The above conclusion becomes further amplified when considering how rarely any such simulated contaminant would end up within the 2.4$^{\prime \prime}$ separation of HIP 79098 (AB)b from the central stellar pair. Given that the simulations yielded 1675 objects across 1 deg$^2$, it follows that the probability of a chance projection of such an object within 2.4$^{\prime \prime}$ separation from HIP 79098 AB (with {any} proper motion) is only 0.2\%. As a double check we  also performed an essentially equivalent procedure on observational 2MASS data. Based on the count of objects that are as bright as or brighter than HIP 79098 (AB)b in the $K$ band in a 15$^{\prime}$ by 15$^{\prime}$ field of view centered on HIP 79098 AB, we derive a probability of 0.3\% that any such object should occur at or within the separation of HIP 79098 (AB)b by chance. Both the Besan\c{c}on simulation and the observational data thus consistently predict a very low chance alignment probability for HIP 79098 (AB)b, irrespective of proper motion.

At the distance of 146.3$\pm$2.5\,pc to the HIP 79098 system (see Sect. \ref{s:binary}), the projected separation of 2.359$\pm$0.001\,$\arcsec$ in the most precise epoch from SPHERE corresponds to 345$\pm$6\,AU for the physical projected separation between the central pair and companion.

\section{Photometric analysis}
\label{s:photo}

\begin{figure*}[htb]
\centering
\setlength{\unitlength}{\textwidth}
\begin{picture}(1.0,0.5)
\put(0.0,0.){\includegraphics[width=0.5\textwidth]{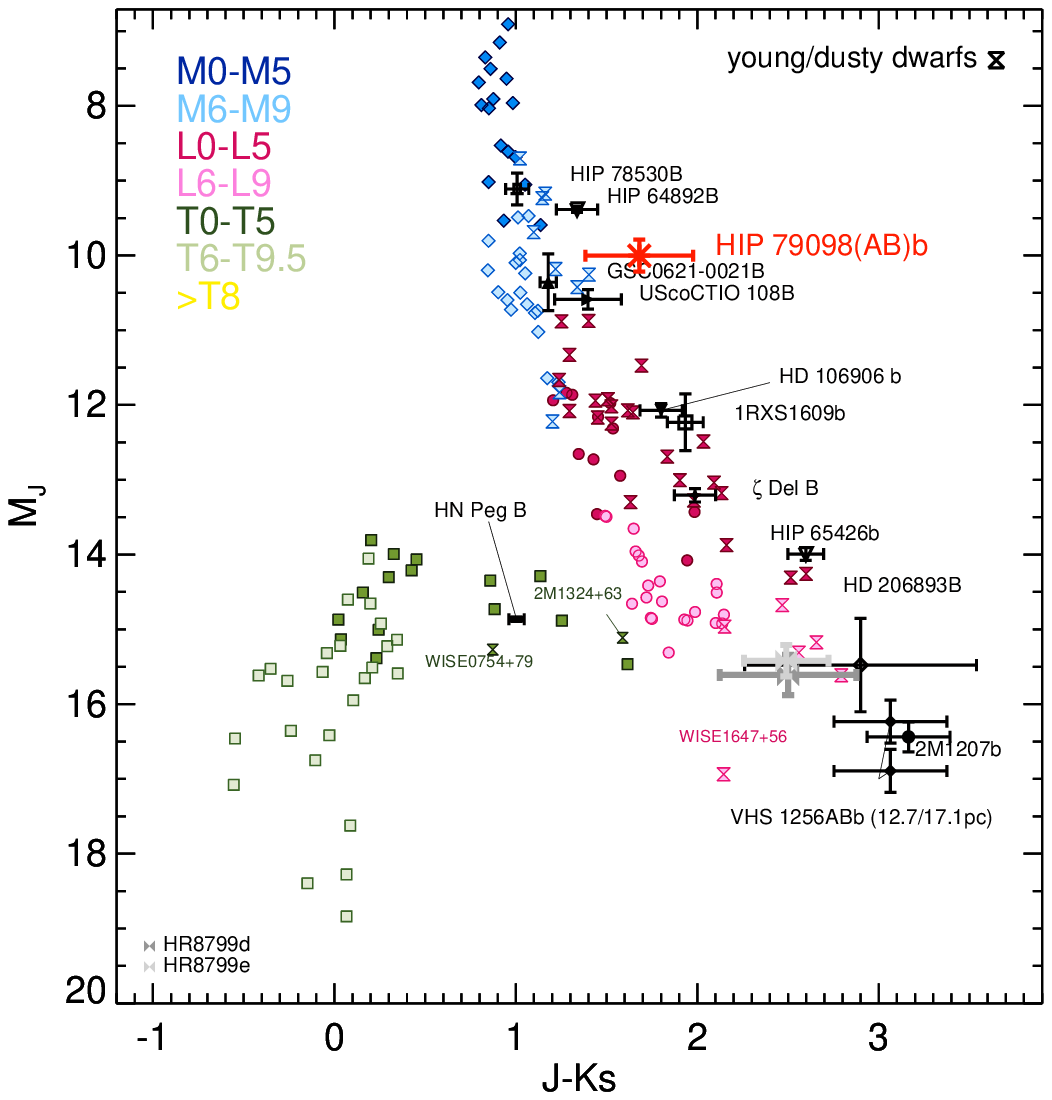}}
\put(0.5,0.){\includegraphics[width=0.5\textwidth]{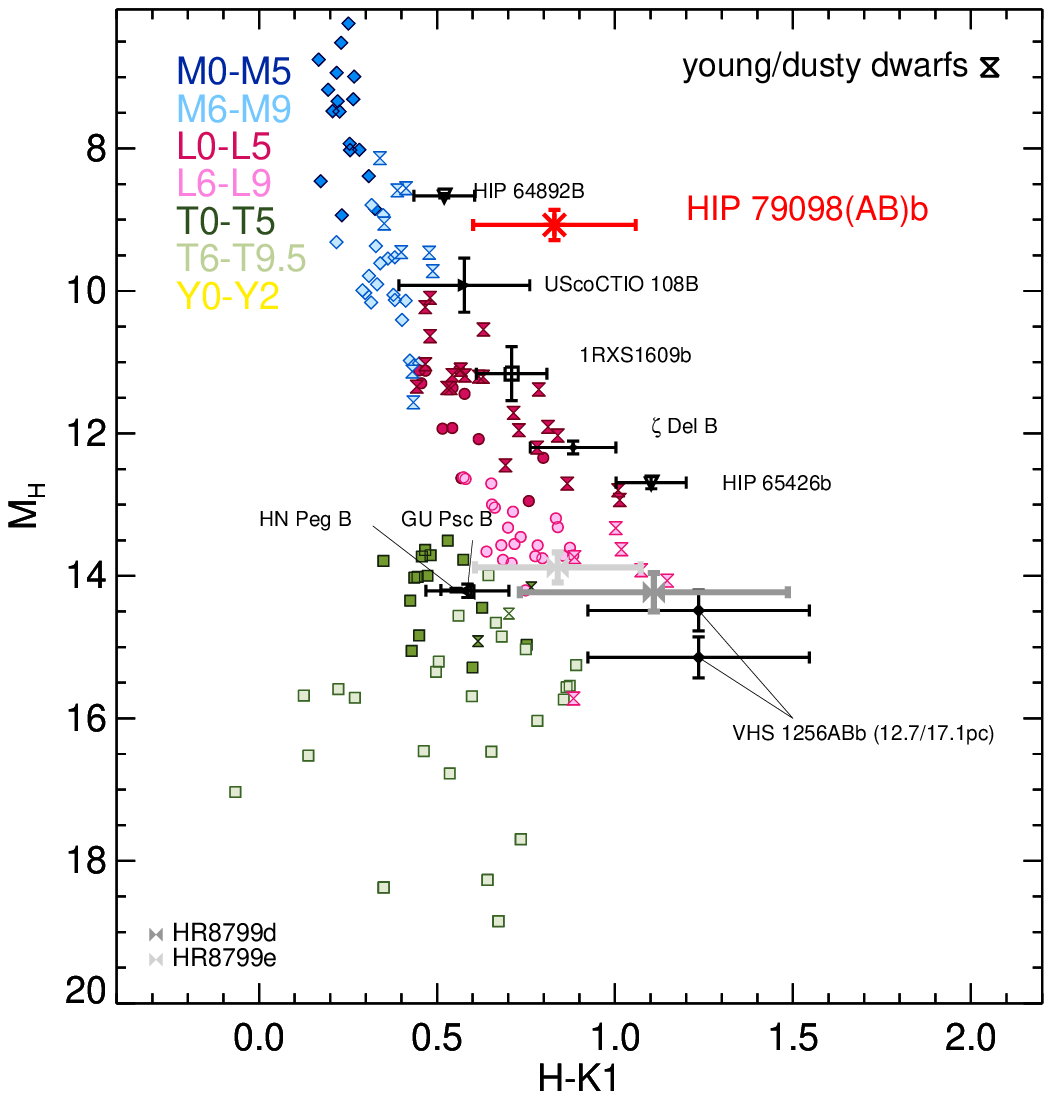}}
\end{picture}
\caption{Color-magnitude diagrams of HIP 79098(AB)b and other substellar objects. Left panel: $M_{J}$ vs. $J-K_{\rm s}$. Right panel: $M_H$ vs. $H-K_1$. The red symbol is HIP 79098 (AB)b. Black symbols are known young objects, while colored symbols are field brown dwarfs. HIP 79098 (AB)b is on the red side of the field L-type sequence, consistently with other young low-gravity objects, and clearly distinct from older field objects.}
\label{f:phot}
\end{figure*}

The photometric values we  derived are shown in Table \ref{t:photometry}. All of these values are consistent (within the error bars) with the literature values in ST02 and K07, with the exception of the $H$ value which is 0.76 mag (3.3$\sigma$) fainter in our analysis compared to the K07 value based on the same data set. We  double-checked  our values and cross-checked our procedures within our team, and did not identify any reason to expect an uncertainty beyond the error bars we  derived. We note that the difference corresponds almost exactly to a factor of 2 in flux, which could potentially reflect differences, for example in  the normalization of direct integration time, which is 0.5 s in $H$ versus 1.0 s in $K_{\rm s}$ for the ND filtered frames. We  checked to verify that we used correct normalizations for each photometric band. However,  since we could not reproduce the K07 value, we simply adopted our own derived value to represent the $H$ magnitude of HIP 79098 (AB)b. Representative color-magnitude diagrams are plotted in Fig. \ref{f:phot}.

\begin{table}[htb]
\caption{Photometric data of HIP 79098 (AB)b.}
\label{t:photometry}
\centering
\begin{tabular}{llll}
\hline
\hline
Band & Facility \& epoch & App. mag & Abs. mag \\
\hline
$J$      & NACO, 2004  & 15.83$\pm$0.21 & 10.00$\pm$0.21 \\
$H$      & NACO, 2004  & 14.90$\pm$0.21 & 9.07$\pm$0.21 \\
$K_{\rm s}$      & NACO, 2004 & 14.15$\pm$0.21 & 8.32$\pm$0.21 \\
$K_1$      & SPHERE, 2015 & 14.07$\pm$0.09 & 8.24$\pm$0.09 \\
$K_2$      & SPHERE, 2015 & 13.85$\pm$0.10 & 8.02$\pm$0.10 \\
\hline
\end{tabular}
\end{table}

As we note  in Sect. \ref{s:intro}, young planets and brown dwarfs in the vicinity of the L-type spectral range typically show considerably redder colors (by $\sim$0.1-0.5 mag in $J-K$) than their older and more massive counterparts of the same SpT. HIP 79098 (AB)b unambiguously displays this trend in both our color-magnitude diagrams, underlining the fact that it must be a young low-gravity object, which is consistent  with what the CPM analysis implies. This trend also naturally explains why the companion differs from conventional model expectations, which was the basis for its classification as a background object in previous studies. ST02 reasoned that red candidates in their sample might be caused by heavily reddened background stars. In the case of HIP 79098 (AB)b this can be excluded on the basis of the CPM analysis, but more generally, we also note that interstellar extinction to the required level should be very unusual. As an illustrative example, we can consider that an extinction of approximately $E(H-K) > 0.35$ would be required to start reproducing the colors of HIP 79098 (AB)b within the error bars, even for very late-type background stars. The predicted extinction levels for background stars are much lower than this. For example, \citet{schlafly2011} give a maximum possible $E(B-V)$ of 0.157 mag in the direction of HIP 79098. This corresponds to an $E(H-K)$ of 0.03 mag. Thus, the interstellar medium cannot produce extinction to the required level. Substantial amounts of circumstellar extinction would be necessary, which would be highly unusual in any representative population of background stars.

\begin{figure}
\setlength{\unitlength}{\textwidth}
\hspace*{-0.5cm}                                                           
 \includegraphics[width=0.55\textwidth]{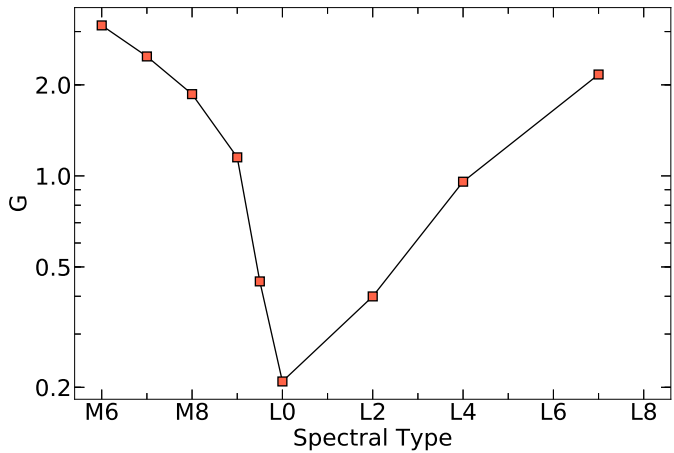}
\caption{G goodness-of-fit statistic for HIP 79098 (AB)b as a function of spectral type for the \citep{luhman2017} templates constructed from a combination of USco and TWA objects. }
\label{f:gval}
\end{figure}

\begin{figure}
\setlength{\unitlength}{\textwidth}
\hspace*{-0.5cm}                                                           
 \includegraphics[width=0.5\textwidth]{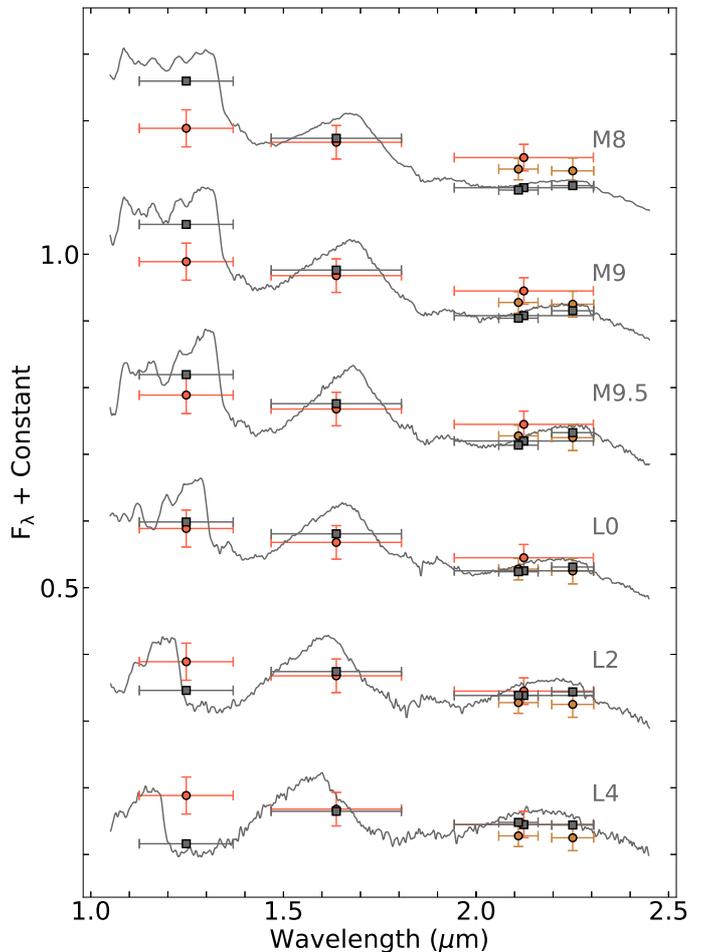}

\caption{Comparison of the observed photometric data of HIP 79098 (AB)b to \citep{luhman2017} templates of different spectral types. NACO $JHK\rm_{s}$ and SPHERE K12 values are shown as red and brown circles, respectively. The grey squares are the result of applying the different filters' transmission curves to the Luhman models (grey curves). Error bars in the x direction correspond to the FWHM of each corresponding filter.}
\label{f:luhm_SpT}
\end{figure}

Using the photometric data points listed in Table \ref{t:photometry}, we can attempt to estimate a spectral type for HIP 79098 (AB)b. To this end, we adopted the \citet{luhman2017} near-IR standard spectral templates for classifying young brown dwarfs and low-mass stars. These templates were constructed from the combination of several objects per spectral type bin in the M-L spectral type range. Here we use the older population (older than a few Myr), which was compiled from several USco and TW Hya members; since the estimated age is $\sim$\,10\,Myr for both regions \citep{pecaut2016, bell2015}, these templates should present  spectral features similar to those of  HIP 79098 (AB)b. We use the G goodness-of-fit statistic, which accounts for the relative width of the various filters, to fit the templates to the data \citep{cushing2008}. The results are presented in Fig. \ref{f:gval}, where the best-fit model seems to be centered around L0. From this analysis, it seems reasonable to set a conservative good-fit range whenever G is below 1.4, which translates to spectral types within the M9--L4 domain. The comparison of these spectral templates to the HIP 79098 (AB)b photometric values is shown in Fig. \ref{f:luhm_SpT}. Given the limited resolution of our data, it is challenging to set a stringent confidence level on the spectral type. However, the near-IR spectrum of HIP 79098 (AB)b appears to be fairly flat, which discards early and medium M types as these objects present a  steeper slope towards longer wavelengths. Medium to later L types are likewise not probable as they become too faint in the $J$ band. We thus deduce that the best-fit spectral type lies between late M-type and early L-type objects.

\begin{figure*}
\centering
\setlength{\unitlength}{\textwidth}
\begin{picture}(1,0.4)
  \put(0.,0.){\includegraphics[width=0.35\textwidth]{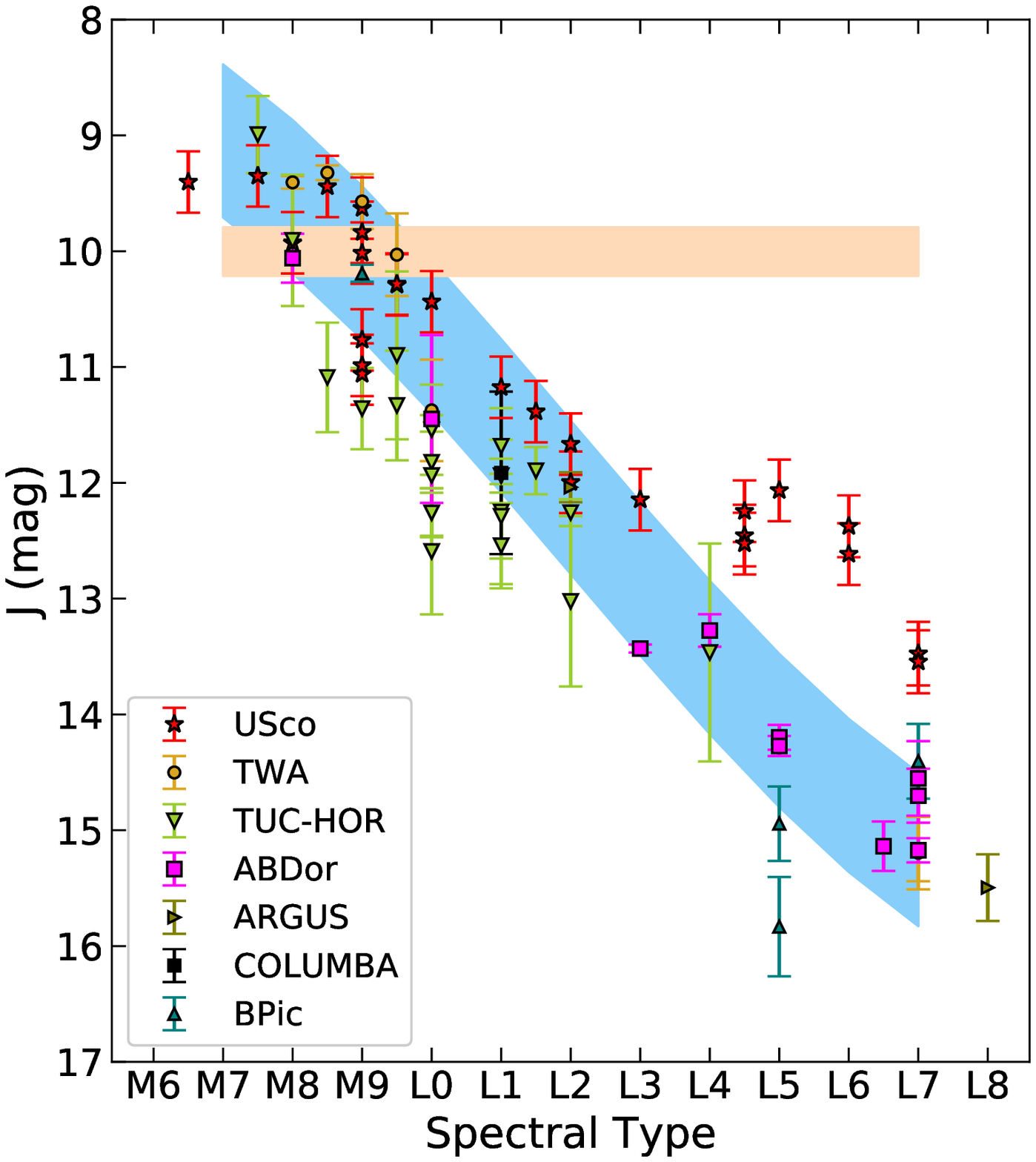}}
  \put(0.33,0.){\includegraphics[width=0.35\textwidth]{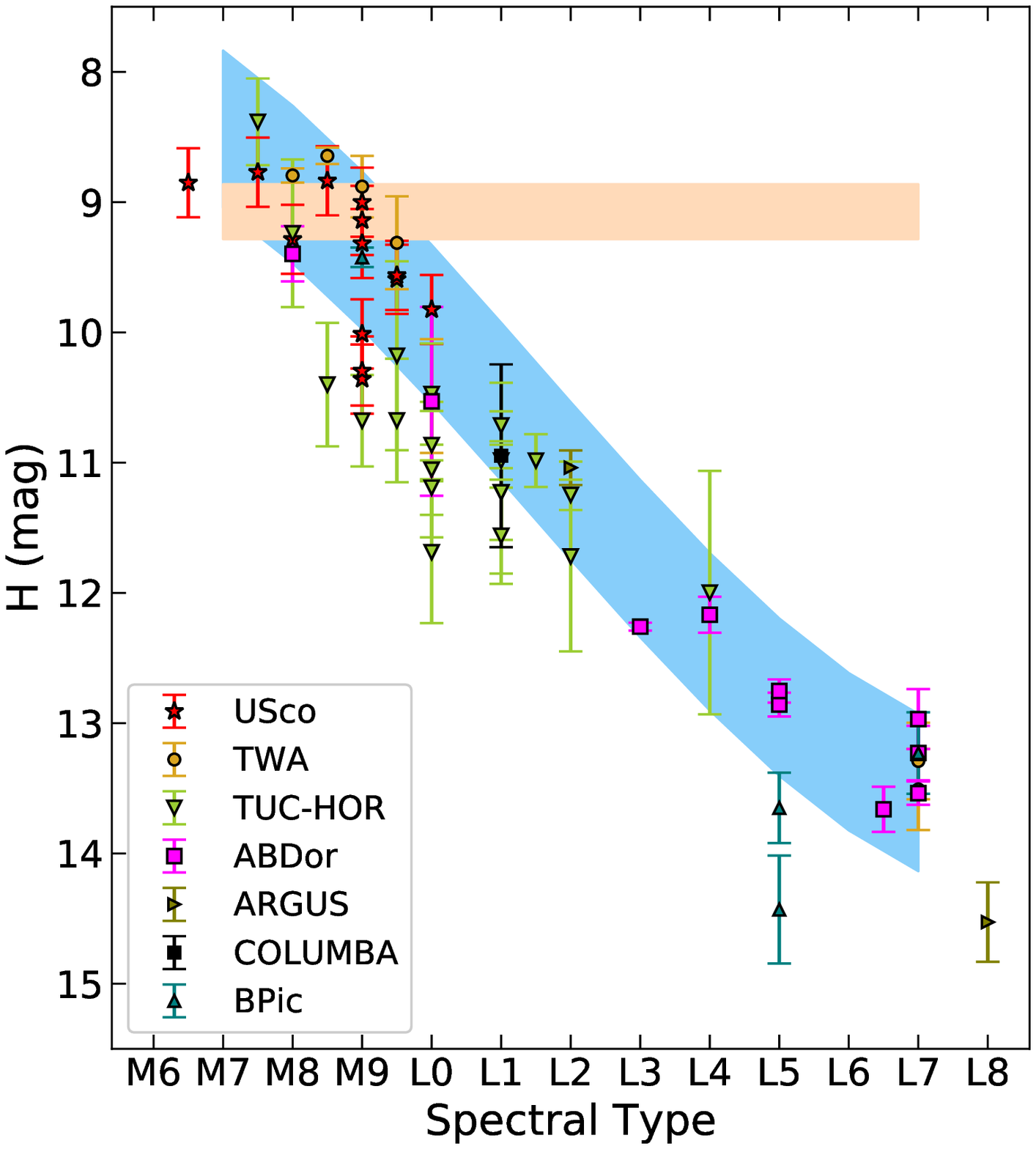}}
 \put(0.66,0.0){\includegraphics[width=0.35\textwidth]{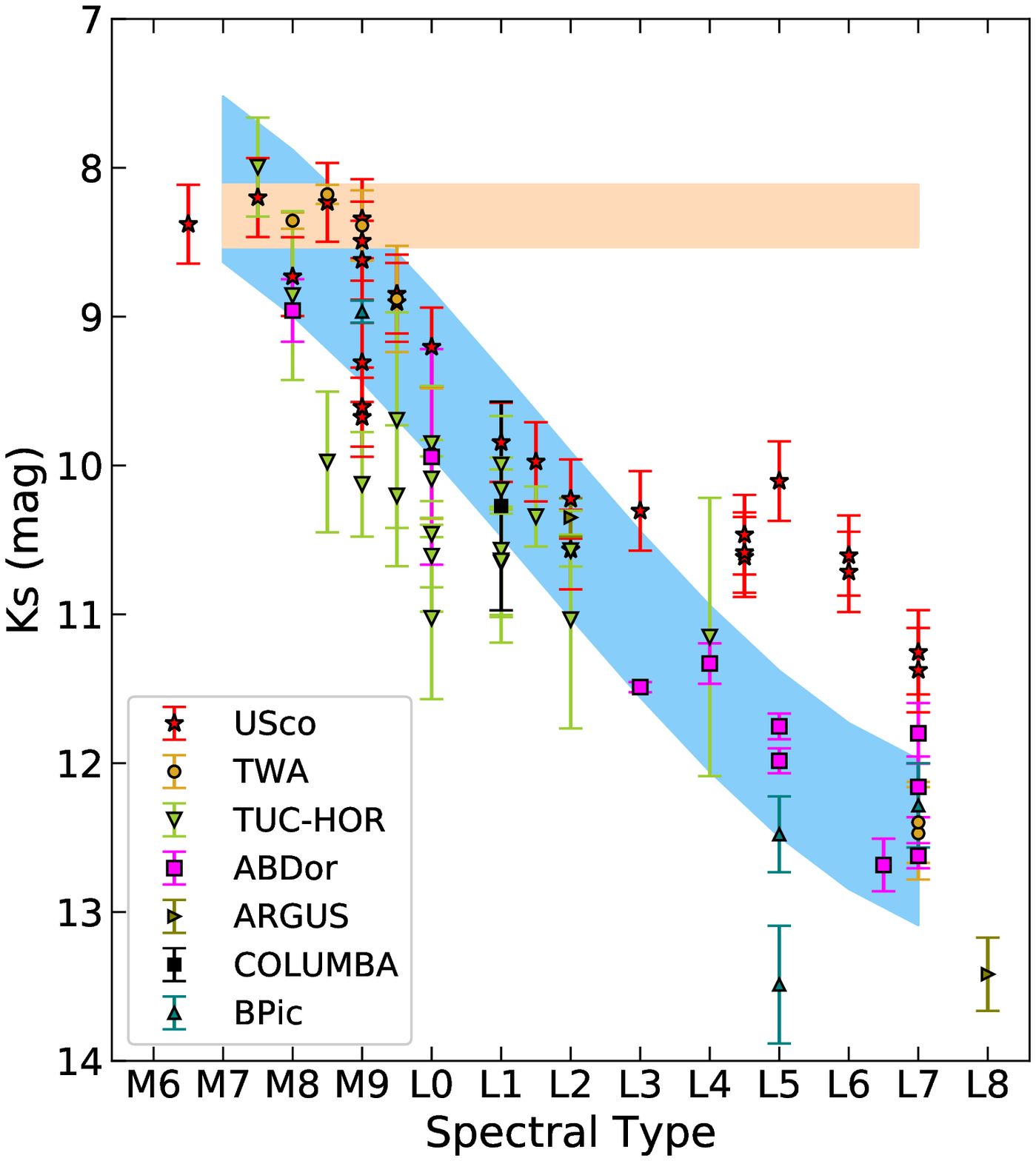}}

\end{picture}
\caption{Absolute magnitude of young low-gravity objects as a function of spectral type (see text for details). The blue-shaded area corresponds to the \citet{faherty2016} polynomial relation for M7--L7 YMG objects, while the horizontal area in orange indicates the magnitude within error bars of HIP 79098 (AB)b for each band.  }
\label{fig:phot_SpT}
\end{figure*}

To further narrow down the list of possible spectral types, we compared the absolute magnitude of HIP 79098 (AB)b with archival objects members of USco and young moving groups (YMGs). We collected the high-confidence low-mass YMG objects presented in \citet{faherty2016} and their best-fit polynomial that accounts for absolute magnitude variation with spectral type. This data set was complemented with  six late L-type objects discovered by \citet{schneider2017} with a YMG membership probability higher than 75\,$\%,$ as computed by the BANYAN II tool \citep{gagne2014}, and with several very low-mass members of USco confirmed by \citet{lodieu2018} and references therein. These diagrams are shown in Fig. \ref{fig:phot_SpT}. The photometric values of HIP 79098 (AB)b place it well above the mid- and late L-type objects, as there seems to be an abrupt brightness transition from M- to L-types in young low-gravity objects.  With a brightness in $JHK\rm{s}$ comparable to young objects of late M spectral type, it seems unlikely that  HIP 79098 (AB)b could be classified as a L2 (or later) type, even accounting for the presence of circumstellar material. From the combination of the fit to the spectral templates and this photometric comparison to young objects, we thus conclude that the most representative spectral type of HIP 79098 (AB)b lies in the M9--L0 range, which also agrees well with the color-magnitude diagrams presented in Fig. \ref{f:phot}. A spectroscopic study of this brown dwarf would help to further constrain its spectral type and other atmospheric and physical properties.

This result can be compared to the Sco-Cen brown dwarf companion HIP 64892 B found by SPHERE during the SHINE campaign \citep{cheetham2018}. As is  discussed further in Sect. \ref{s:discussion}, the HIP 64892 system closely resembles HIP 79098 in many regards, including very similar absolute magnitudes of the brown dwarf companions in the $K$-band range, although HIP 79098 (AB)b is fainter at shorter wavelengths (see Fig. \ref{f:phot}) and therefore is probably a bit colder. Complementing photometric observations with SPHERE long-slit spectroscopy and archival NACO $L$-band data, \citet{cheetham2018} find a best-fit spectral type for HIP 64892 b of M9, corresponding to a T$\rm_{eff}$ of 2600\,K, which agrees well with the upper bounds of our spectral type range for HIP 79098 (AB)b.

\section{Discussion}
\label{s:discussion}

Using BT-SETTL tracks \citep{baraffe2015} to model HIP 79098 (AB)b and assuming the mean USco age of 10 Myr \citep{pecaut2016} for the individual photometric bands, we get masses of 16\,$M_{\rm Jup}$ in $J$, 18\,$M_{\rm Jup}$ in $H$, 20\,$M_{\rm Jup}$ in $K\rm_{s}$, 25\,$M_{\rm Jup}$ in $K_1$, and 23\,$M_{\rm Jup}$ in $K_2$. This corresponds to effective temperatures of 2300\,K in the lowest mass case and 2600\,K in the highest mass case. There is a gradient in mass/$T_{\rm eff}$ with increasing wavelength, which reflects the fact that the companion is redder than the model predictions. This shows  that the BT-SETTL models are not fully applicable to young low-mass objects, as was also seen for  HIP 64892 B \citep{cheetham2018}. Meanwhile, since the $JHK$ range covers a substantial fraction of the energy output of this class of objects, bolometric arguments would imply that the derived temperature range is probably representative of the object. It is certainly consistent with the spectral types derived in the previous section \citep{filippazzo2015}.

A mass range of 16-25\,$M_{\rm Jup}$ corresponds to a mass ratio of 0.6\,\% to 1\,\% for HIP 79098 (AB)b relative to the primary (A) component. Arguably,  a more interesting quantity would be the ratio of the mass of HIP 79098 (AB)b to the total mass of the central AB pair; however, this is  more uncertain since the mass of the B component is unknown (see Sect. \ref{s:binary}). Adopting the full possible span of masses for the AB pair of 2.5-5~$M_{\rm sun}$, we get a total mass ratio of 0.3-1\,\%. Nevertheless, all indications are that the mass ratio of HIP 79098 (AB)b to the central pair is $<$1\,\%. If treated  analogously to  Sun-like stars, this would be on the planetary side of the brown dwarf desert, which is particularly well characterized at small and intermediate separations \citep{grether2006}. HIP 79098 (AB)b joins a growing number of targets in this category around early-type stars.

Two particularly interesting points of comparison in this context are the already mentioned HIP 64892 B and HIP 78530 B \citep{lafreniere2011}. At $\sim$345\,AU, HIP 79098 (AB)b is enveloped in projected separation between HIP 64892 B ($\sim$149 AU) and HIP 78530 B ($\sim$710 AU). All three objects accompany B9/B9.5-type primaries. HIP 78530 is intrinsically a bit brighter than HIP 79098 (AB)b and HIP 64892 B, and also has a somewhat earlier estimated spectral type of M7-M8.5 \citep[][and Petrus et al., in prep.]{lachapelle2015}. It  has the same age as the HIP 79098 system, and an estimated mass of 21-25 $M_{\rm Jup}$, which is consistent with the upper range of our mass estimate for HIP 79098 (AB)b. HIP 79098 (AB)b and HIP 64892 B have the same brightness in the $K$ band range, but HIP 79098 (AB)b is fainter in $J$ and $H$, implying that HIP 79098 (AB)b is redder and colder than HIP 64892 B. It is is also associated with a younger subregion of Sco-Cen than HIP 64892 B, so as expected, its mass estimation is lower than the 29-37 $M_{\rm Jup}$ estimation in \citet{cheetham2018} for HIP 64892 B, although the difference is  somewhat impacted by the different sets of models used\footnote{COND-based tracks \citep{allard2001,baraffe2003} were used for HIP 64892 B}. In contrast to HIP 79098 (AB)b, none of the other systems discussed  has any reported stellar binarity. HD 106906 hosts another circumbinary companion in Sco-Cen \citep{bailey2014}, although  the stellar and companion masses are both substantially lower than in the HIP 79098 system. Statistical surveys have so far shown no significant differences in the substellar companion populations between single and multiple stars \citep{bonavita2016,asensio2018}. HIP 79098 (AB)b appears consistent with this trend. Along with $\kappa$ And b \citep{carson2013}, a population of objects with masses above the classical deuterium burning limit \citep{spiegel2011} but small mass ratios to B-type stars appears to be emerging. This naturally raises the question of whether they may constitute the upper mass end of a planetary population. A coherent statistical survey will  be required  to evaluate this possibility, which is one of the primary purposes of BEAST.

As we have discussed, HIP 79098 (AB)b was classified as a background star in ST02 and K07 based on the fact that it deviated from conventional evolutionary models, whereas we now know that young brown dwarfs in fact do systematically  deviate from those models. This potentially means not only that HIP 79098 (AB)b was misclassified, but also that any other low-mass substellar companion that may have been observed in these studies would probably be systematically classified as background stars as well. This emphasizes the importance of CPM analysis for companionship determination, which in contrast to spectrophotometric fitting is model-free (although a galactic kinematic model can be required to interpret the result in some cases). It also emphasizes the fact that candidates from literature studies that only use photometric criteria to assess physical companionship will need to be followed up and tested for common proper motion. Identifying false negatives (and false positives) is crucial for the statistical interpretation of surveys for wide substellar companions. Based on our results, it is conceivable that the frequency of wide substellar companions may have been underestimated in photometric surveys, particularly for young and massive stars.

\section{Conclusions}
\label{s:summary}

In this paper, we  presented astrometric and photometric evidence that HIP 79098 (AB)b is a young ($\sim$10\,Myr) circumbinary low-mass brown dwarf at a projected separation of $\sim$345\,AU with a model-dependent mass of 16-25 $M_{\rm Jup}$. Two additional co-distant and potentially co-moving wide stellar components (2MASS J16084836-2341209 and 2MASS J16083908-2340055) may exist in the system at 9500 AU and 12900 AU separation respectively, but it is not yet possible to conclude whether they are physically bound to the system. Given a central binary mass of 2.5-5~$M_{\rm sun}$, the estimated substellar companion mass implies a mass ratio to the central binary of 0.3-1\%, which is in the same range as the population of wide directly imaged planets around Sun-like and intermediate-mass stars. Future systematic studies may reveal whether this recently discovered and growing population of objects share a common formation path, and whether this path is in turn the same as or distinct from the population of closer-in planets and low-mass brown dwarfs discovered in RV and transit studies.

\begin{acknowledgements}
M.J. gratefully acknowledges funding from the Knut and Alice Wallenberg Foundation. This study made use of the CDS services SIMBAD and VizieR, and the SAO/NASA ADS service. R.G. and S.D. are supported by the project PRIN-INAF 2016 The Cradle of Life - GENESIS-SKA (General Conditions in Early Planetary Systems for the rise of life with SKA). We also acknowledge support from INAF/Frontiera (Fostering high ResolutiON Technology and Innovation for Exoplanets and Research in Astrophysics) through the ``Progetti Premiali'' funding scheme of the Italian Ministry of Education, University, and Research. A.B. acknowledges support from the European Research Council under ERC Starting Grant agreement 678194 (FALCONER). J.C.C. acknowledges support from SC Space Grant. G.-D.M. acknowledges the support of the DFG priority program SPP 1992 ``Exploring the Diversity of Extrasolar Planets'' (KU 2849/7-1). Part of this research was carried out at the Jet Propulsion Laboratory, California Institute of Technology, under a contract with the National Aeronautics and Space Administration (NASA). E.E.M. acknowledges support from the Jet Propulsion Laboratory Exoplanetary Science Initiative and the NASA NExSS Program. This work has made use of the SPHERE Data Centre, jointly operated by OSUG/IPAG (Grenoble), PYTHEAS/LAM/CeSAM (Marseille), OCA/Lagrange (Nice), Observatoire de Paris/LESIA (Paris), and Observatoire de Lyon/CRAL, and is supported by a grant from Labex OSUG@2020 (Investissements d’avenir – ANR10 LABX56).
\end{acknowledgements}


\begin{thebibliography}{}

\bibitem[Abt \& Morell(1995)]{abt1995} Abt, H.A. \& Morrell, N.I.\ 1995, ApJS, 99, 135
\bibitem[Allard et al.(2001)]{allard2001} Allard, F., Hauschildt, P., Alexander, D.R., Tamanai, A., \& Schweitzer, A.\ 2001, ApJ, 556, 357
\bibitem[Asensio-Torres et al.(2018)]{asensio2018} Asensio-Torres, R., Janson, M., Bonavita, M. et al.\ 2018, A\&A, 619, 43
\bibitem[Bailey et al.(2014)]{bailey2014} Bailey, V., Meshkat, T., Reiter, M. et al.\ 2014, ApJ, 780, L4
\bibitem[Baraffe et al.(1998)]{baraffe1998} Baraffe, I., Chabrier, G., Allard, F., \& Hauschildt, P.H. 1998, A\&A, 337, 403
\bibitem[Baraffe et al.(2003)]{baraffe2003} Baraffe, I., Chabrier, G., Barman, T.S., Allard, F., \& Hauschildt, P.\ 2003, A\&A, 402, 701
\bibitem[Baraffe et al.(2015)]{baraffe2015} Baraffe, I., Homeier, D., Allard, F. et al.\ 2015, A\&A, 577, 42
\bibitem[Becker et al.(2015)]{becker2015} Becker, J.C., Johnson, J.A., Vanderburg, A., \& Morton, T.D.\ 2015, ApJS, 217, 29
\bibitem[Beuzit et al.(2019)]{beuzit2019} Beuzit, J.-L., Vigan, A., Mouillet, D. et al.\ 2019, A\&A, submitted, arXiv:1902.04080
\bibitem[Bell et al.(2015)]{bell2015} Bell, C. P. M., Mamajek, E. E., Naylor, T. 2015, MNRAS, 454, 593
\bibitem[Bertiau(1958)]{bertiau1958} Bertiau, F.C. 1958, ApJ, 128, 533
\bibitem[Bonavita et al.(2016)]{bonavita2016} Bonavita, M., Desidera, S., Thalmann, C., Janson, M., Vigan, A., Chauvin, G., \& Lannier, J. 2016, A\&A, 593, 38
\bibitem[Bonnefoy et al.(2016)]{bonnefoy2016} Bonnefoy, M., Zurlo, A., Baudino, J.L. et al. 2016, A\&A, 587, 58
\bibitem[Brandt et al.(2014)]{brandt2014} Brandt, T.D., Kuzuhara, M., McElwain, M.W. et al. 2014, ApJ, 786, 1
\bibitem[Brown \& Verschueren(1997)]{brown1997} Brown, A.G.A. \& Verschueren, W.\ 1997, A\&A, 319, 811
\bibitem[Brown et al.(2018)]{brown2018} Brown, A.G.A., Vallenari, A. Prusti, T. et al.\ 2018, A\&A, 616, 1 
\bibitem[Carson et al.(2013)]{carson2013} Carson, J., Thalmann, C., Janson, M. et al.\ 2013, ApJ, 763, L32
\bibitem[Castelli(1991)]{castelli1991} Castelli, F.\ 1991, A\&A, 251, 106
\bibitem[Chabrier et al.(2000)]{chabrier2000} Chabrier, G., Baraffe, I., Allard, F., \& Hauschildt, P. 2000, ApJ, 542, 464
\bibitem[Chauvin et al.(2017)]{chauvin2017} Chauvin, G., Desidera, S., Lagrange, A.-M. et al.\ 2017, A\&A, 605, L9
\bibitem[Cheetham et al.(2018)]{cheetham2018} Cheetham, A., Bonnefoy, M., Desidera, S. et al.\ 2018, A\&A, 615, 160
\bibitem[Cushing et al.(2008)]{cushing2008} Cushing, M. C., Marley, M. S., Saumon, D., et al. 2008, ApJ, 678, 1372
\bibitem[Delorme et al.(2017)]{delorme2017} Delorme, P., Meunier, N., Albert, D. et al.\ 2017, in SF2A-2017: Proceedings of the Annual meeting of the French Society of Astronomy and Astrophysics, ed. C. Reyl{\`e}, P. Di Matteo, F. Herpin, E. Lagadec, A. Lancon, Z. Meliani, \& F. Royer, 347-361
\bibitem[de Zeeuw et al.(1999)]{dezeeuw1999} de Zeeuw, P.T., Hoogerwerf, R., de Bruijne, J.H.J., Brown, A.G.A., \& Blaauw, A.\ 1999, AJ, 117, 354
\bibitem[Faherty et al.(2016)]{faherty2016} Faherty, J.K., Riedel, A.R., Cruz, K.L. et al.\ 2016, ApJS, 225, 10 
\bibitem[Filippazzo et al.(2015)]{filippazzo2015} Filippazzo, J.C., Rice, E.L., Faherty, J.K., Cruz, K.L., van Gordon, M.M., \& Looper, D.L.\ 2015, ApJ, 810, 158
\bibitem[Fiorucci \& Munari(2003)]{fiorucci2003} Fiorucci, M. \& Munari, U.\ 2003, A\&A, 401, 781
\bibitem[Gagn{\'e} et al.(2014)]{gagne2014} Gagne, J.,Lafreniere, D., Doyon, R. et al.\ 2004, ApJ, 783, 121
\bibitem[Gagn{\'e} et al.(2018)]{gagne2018} Gagne, J., Mamajek, E., Malo, L. et al.\ 2018, ApJ, 856, 23
\bibitem[Galicher et al.(2018)]{galicher2018} Galicher, R., Boccaletti, A., Mesa, D. et al.\ 2018, A\&A, 615, 92
\bibitem[Gizis et al.(2015)]{gizis2015} Gizis, J.E., Allers, K., Liu, M.C., Harris, H.C., Faherty, J.K., Burgasser, A.J., \& Kirkpatrick, J.D.\ 2015, ApJ, 799, 203
\bibitem[Grether et al.(2006)]{grether2006} Grether, D, \& Lineweaver, C.\ 2006, ApJ, 640, 1051
\bibitem[Hartoog(1977)]{hartoog1977} Hartoog, M.R.\ 1977, ApJ, 212, 723 
\bibitem[Huber et al.(2016)]{huber2016} Huber, D., Bryson, S.T., Haas, M.R. et al.\ 2016, ApJS, 224, 2 
\bibitem[Ireland et al.(2011)]{ireland2011} Ireland, M., Kraus, A., Martinache, F., Law, N., \& Hillenbrand, L.A.\ 2011, ApJ, 726, 113
\bibitem[Janson et al.(2011)]{janson2011} Janson, M., Bonavita, M., Klahr, H., Lafreni{\`e}re, D., Jayawardhana, R., Zinnecker, H.\ 2011, ApJ, 736, 89
\bibitem[Janson et al.(2012)]{janson2012} Janson, M., Carson, J., Lafreni{\`e}re, D. et al.\ 2012, ApJ, 758, L2
\bibitem[Janson et al.(2013)]{janson2013} Janson, M., Lafreni{\`e}re, D., Jayawardhana, R., Bonavita, M., Girard, J., Brandeker, A., \& Gizis, J.E. 2013, ApJ 773, 170
\bibitem[Keppler et al.(2018)]{keppler2018} Keppler, M., Benisty, M., M{\"u}ller, A. et al.\ 2018, A\&A, 617, 44
\bibitem[Kouwenhoven et al.(2005)]{kouwenhoven2005} Kouwenhoven, M.B.N., Brown, A.G.A., Zinnecker, H., Kaper, L., \& Portegies Zwart, S.F. 2005, A\&A, 430, 137
\bibitem[Kouwenhoven et al.(2007)]{kouwenhoven2007} Kouwenhoven, M.B.N., Brown, A.G.A., \& Kaper, L. 2007, A\&A, 464, 581
\bibitem[Lachapelle et al.(2015)]{lachapelle2015} Lachapelle, F.-R., Lafreni{\`e}re, D., Gagn{\'e}, J., Jayawardhana, R., Janson, M., Helling, C., \& Witte, S.\ 2015, ApJ, 802, 61
\bibitem[Lafreni{\`e}re et al.(2009)]{lafreniere2009} Lafreni{\`e}re, D., Marois, C., Doyon, R., \& Barman, T.\ 2009, ApJ, 694, L148
\bibitem[Lafreni{\`e}re et al.(2011)]{lafreniere2011} Lafreni{\`e}re, D., Jayawardhana, R., Janson, M., Helling, C., Witte, S. \& Hauschildt, P.\ 2011, ApJ, 730, 42
\bibitem[Lafreni{\`e}re et al.(2014)]{lafreniere2014} Lafreni{\`e}re, D., Jayawardhana, R., van Kerkwijk, M.H., Brandeker, A., \& Janson, M.\ 2014, ApJ, 785, 47
\bibitem[Lagrange et al.(2010)]{lagrange2010} Lagrange, A.-M., Bonnefoy, M., Chauvin, G. et al.\ 2010, Science, 329, 57
\bibitem[Lenzen et al.(2003)]{lenzen2003} Lenzen, R., Hartung, M., Brandner, W. et al.\ 2003, SPIE, 4841, 944 
\bibitem[Levato et al.(1987)]{levato1987} Levato, H., Malaroda, S., Morrell, N., \& Solivella, G.\ 1987, ApJS, 64, 487
\bibitem[Liu et al.(2013)]{liu2013} Liu, M.C., Magnier, E.A., Deacon, N.R. et al.\ 2013, ApJ, 777, L20
\bibitem[Lodieu et al.(2007)]{lodieu2007} Lodieu, N., Hambly, N.C., Jameson, R.F., Hodgkin, S.T., Carraro, G., \& Kendall, T.R.\ 2007, MNRAS, 374, 372
\bibitem[Lodieu et al.(2011)]{lodieu2011} Lodieu, N., Dobbie, P.D., \& Hambly, N.C.\ 2011, A\&A, 527, 24
\bibitem[Lodieu et al.(2018)]{lodieu2018} Lodieu, N., Zapatero Osorio, M.R.; B{\'e}jar, V.J.S. et al.\ 2018, MNRAS, 473, 2020 
\bibitem[Luhman \& Mamajek(2012)]{luhman2012} Luhman, K.L. \& Mamajek, E.E.\ 2012, ApJ, 758, 31
\bibitem[Luhman et al.(2017)]{luhman2017} Luhman, K.L., Mamajek, E.E., Shukla, S.J. et al.\ 2017, AJ, 153, 46
\bibitem[Macintosh et al.(2015)]{macintosh2015} Macintosh, B., Graham, J.R., Barman, T. et al.\ 2015, Science, 350, 64
\bibitem[Marois et al.(2008)]{marois2008} Marois, C., Macintosh, B., Barman, T., Zuckerman, B., Song, I., Patience, J., Lafreni{\`e}re, D., \& Doyon, R\ 2008, Science, 322, 1348
\bibitem[Mason et al.(2001)]{mason2001} Mason, B.D., Wycoff, G.L., Hartkopf, W.I., Douglass, G.G., \& Worley, C.E. 2001, AJ 122, 3466
\bibitem[Neuh{\"a}user et al.(2005)]{neuhauser2005} Neuh{\"a}user, R., Guenther, E.W., Wuchterl, G., Mugrauer, M., Bedalov, A., \& Hauschildt, P.\ 2005, A\&A, 435, L13
\bibitem[Nielsen et al.(2017)]{nielsen2017} Nielsen, E.L., De Rosa, R.J., Rameau, J. et al.\ 2017, AJ, 154, 218
\bibitem[Norris et al.(1971)]{norris1971} Norris, J.\ 1971, ApJS, 23, 213
\bibitem[Pecaut \& Mamajek(2013)]{pecaut2013} Pecaut, M. \& Mamajek, E.E.\ 2013, ApJS, 208, 9
\bibitem[Pecaut \& Mamajek(2016)]{pecaut2016} Pecaut, M. \& Mamajek, E.E.\ 2016, MNRAS, 461, 794
\bibitem[Preibisch et al.(2002)]{preibisch2002} Preibisch, T., Brown, A.G.A., Bridges, T., Guenther, E., \& Zinnecker, H.\ 2002, AJ, 124, 404
\bibitem[Rameau et al.(2013)]{rameau2013} Rameau, J., Chauvin, G., Lagrange, A.-M. et al.\ 2013, ApJ, 772, L15
\bibitem[Riaz et al.(2012)]{riaz2012} Riaz, B., Lodieu, N., Goodwin, S., Stamatellos, D., \& Thompson, M.\ 2012, A\&A, 527, 24
\bibitem[Robin et al.(2003)]{robin2003} Robin, A.C., Reyl{\'e}, C., Derri{\`e}re, S., \& Picaud, S.\ 2003, A\&A, 409, 523 
\bibitem[Rousset et al.(2003)]{rousset2003} Rousset, G., Lacombe, F., Puget, P. et al. 2003, SPIE, 4839, 140 
\bibitem[Schlafly et al.(2011)]{schlafly2011} Schlafly, E.F. \& Finkbeiner, D.P.\ 2011, ApJ, 737, 103
\bibitem[Schneider et al.(1981)]{schneider1981} Schneider, H.\ 1981, A\&AS, 44, 137
\bibitem[Schneider et al.(2017)]{schneider2017} Schneider, A.C., Windsor, J., Cushing, M.C. et al.\ 2017, AJ, 153, 196
\bibitem[Shatsky \& Tokovinin(2002)]{shatsky2002} Shatsky, N. \& Tokovinin, A. 2002, A\&A, 382, 92
\bibitem[Spiegel et al.(2011)]{spiegel2011} Spiegel, D.S., Burrows, A., \& Milson, J.A.\ 2011, ApJ, 727, 57
\bibitem[Stauffer et al.(2018)]{stauffer2018} Stauffer, J., Rebull, L., David, T.J. et al.\ 2018, AJ, 155, 63 
\bibitem[Vigan et al.(2017)]{vigan2017} Vigan, A., Bonavita, M., Biller, B., et al.\ 2017, A\&A, 603, 3 
\bibitem[Wagner et al.(2016)]{wagner2016} Wagner, K., Apai, D., Kasper, M. et al.\ 2016, Science, 353, 673 
\bibitem[Worley et al.(2012)]{worley2012} Worley, C.C., de Laverny, P., Recio-Blanco, A., Hill, V., Bijaoui, A., \& Ordenovic, C.\ 2012, A\&A, 542, 48 

\end{thebibliography}
\end{document}